\documentclass[journal,twoside]{asmb}

\begin{document}
\title{Effects of Co-channel Interference on RIS Empowered Wireless Networks amid Multiple Eavesdropping Attempts}

\author[1]{Md. Roisul Ajom Ruku}
\author[2]{Md. Ibrahim}
\author[3]{A. S. M. Badrudduza}
\author[4]{Imran Shafique Ansari}

\affil[1]{Department of Electrical \& Electronic Engineering,  Rajshahi University of Engineering \& Technology (RUET), Rajshahi-6204, Bangladesh}
\affil[2]{Institute of Information and Communication Technology, RUET}
\affil[3]{Department of Electronics \& Telecommunication Engineering,RUET}
\affil[4]{James Watt School of Engineering, University of Glasgow, Glasgow G12 8QQ, United Kingdom}

\twocolumn[
\begin{@twocolumnfalse}
\maketitle
%&&&&&&&&&&&&&&&<ABSTRACT>&&&&&&&&&&&&&&
\begin{abstract}
\section*{Abstract}

This letter is concerned with the secrecy performance of reconfigurable intelligent surfaces (RIS)-aided wireless networks in the existence of multiple interferers towards the destination. To be more precise, we analyze three critical issues in the design of secure RIS-assisted networks: 1) How do interferers affect the performance of secure wireless networks? 2) Which of the two groups of eavesdroppers (i.e., colluding and non-colluding) is more severe? 3) How can RIS improve network confidentiality? To do so, we develop the analytical expression of secrecy outage probability in closed-form, along with asymptotic analysis at high signal-to-noise ratio regime to better understand the impacts of different system parameters on secrecy performance. Finally, we validate our analytical results using a computer based Monte-Carlo simulation.
\end{abstract}
%&&&&&&&&&&&&&<END-ABSTRACT>&&&&&&&&&&&&
%$$$$$$$$$$$$$$$<KEYWORDs>$$$$$$$$$$$$$$
\begin{IEEEkeywords}
\section*{Keywords} 

Reconfigurable intelligent surface (RIS), co-channel interference (CCI), secrecy outage probability (SOP), eavesdropper
\end{IEEEkeywords}
%$$$$$$$$$$$$$<END-KEYWORD>$$$$$$$$$$$$$
\end{@twocolumnfalse}
]

%%%%%%%%%%%%%%%%<SECTION>%%%%%%%%%%%%%%%
\section{Introduction}

%#############<SUB-SECTION>#############
\subsection{Background and Literature Study}
Reconfigurable intelligent surfaces (RIS) have become a popular topic in sixth-generation ($6$G) wireless networks due to environmental and economic considerations, as they can increase coverage, enhance energy efficiency, and reduce costs \cite{9917328}. In general, RIS consists of numerous passive elements that form a flat meta surface, and these components can be controlled to optimize the phase and amplitude of incoming signals, thereby reducing negative effects on wireless networks \cite{wu2019towards}. Unlike other technologies, the signal does not undergo encoding and decoding during transmission, which has generated interest in RIS as a field of research.

In light of the broadcast nature of the wireless channel, information security has emerged as one of the most pressing issues in next-generation communication. However, physical layer security (PLS) analysis, which ensures confidential transmission perfectly, has been proven to be an effective solution in this context \cite{badrudduza2021security}. Despite being a new technology, the PLS of RIS-aided models has been studied in a few articles \cite{yang2020secrecy,wei2022secrecy,yadav2022secrecy,zhang2022secrecy,tang2021novel} only. For example, the authors of \cite{yang2020secrecy} suggested that deploying RIS between source and destination could improve secrecy performance significantly. In \cite{wei2022secrecy}, the authors studied the PLS for RIS-aided networks with multiple eavesdroppers and demonstrated how antenna diversity can facilitate  secrecy performance. Moreover, integrating RIS in non-orthogonal multiple access networks, secrecy behavior was evaluated recently in \cite{zhang2022secrecy,tang2021novel}. Besides, co-channel interference (CCI), particularly in cellular communications, is inevitable due to the spatial reuse of the frequency spectrum in wireless networks \cite{tania2022combined}. Two relay-assisted multi-user scheduling techniques were proposed in \cite{li2022secrecy} for investigating secrecy characteristics in the presence of CCI. A research in \cite{ssettumba2019physical} reported recently that increasing the number of interferers towards the destination drastically affects secrecy performance.

%#############<SUB-SECTION>#############
\subsection{Motivation and Contributions}

According to the aforementioned studies, the existing literature is generally focused on RIS-aided configurations where Nakagami-$m$ distributions are largely taken into account. Moreover, no research has been conducted yet to realize the impact of CCI on the secrecy performance of such model. In contrast to the current literature, we propose a novel RIS-assisted structure under $\alpha$-$\mu$ distributions wherein multiple interferers affect the destination and eavesdroppers are capable of intercepting information directly from the source or via RIS. We also assume two eavesdropping cases i.e. colluding and non-colluding (NC) schemes in our analysis. Therefore, taking into account generalized $\alpha$-$\mu$ fading models provides the benefits of incorporating some previous research \cite{yang2020secrecy,yadav2022secrecy} as part of our work. To summarize, the key contributions of this paper are given below:
\begin{itemize}
    \item Unlike previous works, which mainly focus on Nakagami-$m$ and Rayleigh distributions for RIS-assisted cellular networks, we introduce $\alpha$-$\mu$ fading models in our analysis for the first time. Moreover, we investigate a general PLS framework for such configuration in the presence of CCIs that is novel to the author's best knowledge.
    \item To analyze the secrecy performance, firstly, we derive the cumulative density function (CDF) for the RIS-aided network. Therefore, we develop novel analytical expressions of secrecy outage probability (SOP) applicable in three different scenarios. More precisely, asymptotic secrecy analysis is also provided for the considered scenarios including colluding and NC eavesdropping attacks to provide some insightful information.
    \item Finally, based on these demonstrated expressions, we conclude the discussion by providing some insight into the design of secure RIS-aided wireless networks. Furthermore, the analytical results are also verified with the aid of Monte-Carlo (MC) simulations.
\end{itemize}

%===============<FIGURE>================
%\begin{figure*}[!ht]
%\vspace{0mm}
 %   \centerline{\includegraphics[width=0.9\textwidth,angle =0]{Figures/Final_SystemModel.pdf}}
     %   \vspace{0mm}
 %   \caption{System model of a combined RIS-aided dual-hop RF-UOWC system with source ($\mathcal{T}$), relay ($\mathcal{R}$), user ($\mathcal{U}$), and eavesdropper ($\mathcal{E}$).}
   % \label{fig:1}
%\end{figure*}
%=============<END-FIGURE>==============
\subsection{Organization}

The outline of this paper is designed as follows: Section \ref{system} delineates models of the proposed system and channel in realization, while Section \ref{metrics} demonstrates the analytical expressions of SOP for two different eavesdroppers, as well as their asymptotic expressions. Numerical outcomes and some important observations are provided in Section \ref{results}. Finally, the paper concludes with Section \ref{conclusion}.
%%%%%%%%%%%%%%%%<SECTION>%%%%%%%%%%%%%%%

\section{System Model and Problem Formulation}
\label{system}
%\vspace{-3mm}
In this paper, a RIS-assisted wireless network is considered where the source ($\mathcal{S}$) aims to communicate sensitive information to the receiver ($\mathcal{D}$) with the help of a RIS ($\mathcal{R}$) that is furnished with $\mathcal{N}_{d}$ passive elements while multiple colluding (\textit{Case-I}) and non-colluding (\textit{Case-II}) eavesdroppers ($\mathcal{E}$) attempt to intercept the confidential communication directly (\textit{Scenario-I}) or via another RIS ($\mathcal{I}$) (\textit{Scenario-II}) having $\mathcal{N}_{E}$ reflecting elements or via RIS ($\mathcal{R}$) (\textit{Scenario-III)}. We assume $\mathcal{D}$ is affected by the interfering signals from $\mathcal{M}$ arbitrary number of interferers. Here, each node (e.g. $\mathcal{S}$ and $\mathcal{D}$) is comprised of a single antenna, and all the links e.g. $\mathcal{S}$-$\mathcal{R}_{k}$ ($k=1,2,\ldots, \mathcal{N}_{d}$), $\mathcal{R}_{k}$-$\mathcal{D}$, $\mathcal{S}$-$\mathcal{E}_{q}$ ($q=1,2,\ldots,\mathcal{L}$), $\mathcal{S}$-$\mathcal{I}_{p}$ ($p=1,2,\ldots, \mathcal{N}_{E}$), $\mathcal{I}_{p}$-$\mathcal{E}_{q}$, and $\mathcal{M}_{i}$-$\mathcal{D}$ ($i=1,2,\ldots,\mathcal{M}$) are subjected to $\alpha$-$\mu$ fading distribution with corresponding channel gains of $h_{sk}=\theta_{k}e^{-j\delta_{k}}, h_{kd}=\phi_{k}e^{-j\beta_{k}}, h_{sq}, h_{sp}=\theta_{p}e^{-j\delta_{p}}, h_{pq}=\phi_{pq}e^{-j\beta_{pq}}, h_{id}$, respectively, where $\theta_{k}$, $\phi_{k}$, $\theta_{p}$, $\phi_{pq}$ are the channel amplitudes and $\delta_{k}$, $\beta_{k}$, $\delta_{p}$, $\beta_{pq}$ refer to the phase amplitudes. Let $\mathcal{P}_{s}$ denotes the transmit signal power from the source, $\mathcal{P}_{i}$ is the average power of the transmit signal from $i$th interferer, and $\mathcal{Z}_{d}$ and $\mathcal{Z}_{eq}$ are the noise powers at $\mathcal{D}$ and $q$th $\mathcal{E}$, respectively. Then, the instantaneous signal-to-noise ratios (SNRs) of $\mathcal{M}_{i}-\mathcal{D}$ and $\mathcal{S}-\mathcal{E}_{q}$ links are given by $\gamma_{id}=\bar{\gamma}_{id}|h_{id}|^{2}$, $\gamma_{sq}=\bar{\gamma}_{sq}|h_{sq}|^{2}$, $\gamma_{e,c}=\sum_{q=1}^{\mathcal{L}}\gamma_{sq}$ (\textit{Case-I}), and $\gamma_{e,nc}= \arg \underset{q=1,2,\ldots,\mathcal{L}}{max}\gamma_{sq}$ (\textit{Case-II}), where the average SNR terms are defined as $\bar{\gamma}_{id}=\frac{\mathcal{P}_{i}}{\mathcal{Z}_{d}}$ and $\bar{\gamma}_{sq}=\frac{\mathcal{P}_{s}}{\mathcal{Z}_{eq}}$. Instantaneous SNRs at $\mathcal{D}$ and $\mathcal{E}$ with the RIS links are, respectively, given by $\gamma_{d}=\frac{\mathcal{P}_{s}}{\mathcal{Z}_{d}}\bigl[\sum_{k=0}^{\mathcal{N}_{d}}\theta_{k}\phi_{k}e^{-j(\Theta_{k}-\delta_{k}-\beta_{k})}\bigl]^{2}$ and $\gamma_{eq}=\frac{\mathcal{P}_{s}}{\mathcal{Z}_{eq}}\bigl[\sum_{p=0}^{\mathcal{N}_{E}}\theta_{p}\phi_{pq}e^{-j(\Theta_{pq}-\delta_{p}-\beta_{pq})}\bigl]^{2}$. In order to maximize the SNRs, the optimal values are chosen as $\Theta_{k}=\delta_{k}+\beta_{k}$ and $\Theta_{pq}=\delta_{p}+\beta_{pq}$, and accordingly the corresponding SNRs are given by $\gamma_{d}=\bar{\gamma}_{d}\bigl[\sum_{k=0}^{\mathcal{N}_{d}}\theta_{k}\phi_{k}\bigl]^{2}$ and $\gamma_{eq}=\bar{\gamma}_{eq}\bigl[\sum_{p=0}^{\mathcal{N}_{E}}\theta_{p}\phi_{p}\bigl]^{2}$, and $\gamma_{E,C}=\sum_{q=1}^{\mathcal{L}}\gamma_{eq}$ (\textit{Case-I}) and $\gamma_{E,NC}= \arg \underset{q=1,2,\ldots,\mathcal{L}}{max}\gamma_{eq}$ (\textit{Case-II}), where $\bar{\gamma}_{d}=\frac{\mathcal{P}_{s}}{\mathcal{Z}_{d}}$ and $\bar{\gamma}_{eq}=\frac{\mathcal{P}_{s}}{Z_{eq}}$ are the average SNRs. Since $\mathcal{D}$ receives interference from $\mathcal{M}$ interferers, the SIR at $\mathcal{D}$ is $\gamma_{d,i}=\frac{\mathcal{P}_{s}\bigl[\sum_{k=0}^{\mathcal{N}_{d}}\theta_{k}\phi_{k}\bigl]^{2}}{\sum_{i=1}^{\mathcal{M}}\mathcal{P}_{i}|h_{id}|^{2}}$.

\subsection{PDF and CDF of $\gamma_{d}$}
The probability density function (PDF) and the CDF of $\gamma_{d}$ can be expressed as
%%%%%%%%%%% RIS PDF EQUATION %%%%%%%%%%%
\begin{align}\label{eqn:1}
f_{\gamma_{d}}(\gamma)&=\frac{(\frac{ \beta_{d}}{\bar{\gamma_{d}}})^{\frac{\vartheta_{d}}{2}}}{2\Gamma{(\vartheta_{d})}\zeta_{d}^{\vartheta_{d}}} \gamma^{\frac{\vartheta_d}{2}-1}e^{-\sqrt{\frac{\beta_{d}}{\bar{\gamma_{d}}\zeta_{d}^{2}}\gamma}}, 
\\\label{eqn:2}
F_{\gamma_{d}}(\gamma)=&1-\sum_{\Lambda_{1}=0}^{\vartheta_{d}-1}\frac{1}{\Lambda_{1}!}\left(\sqrt{\frac{\beta_{d}}{\bar{\gamma_{d}}\zeta_{d}^{2}}}\right)^{\Lambda_{1}}
%\\ \times
\gamma^{\frac{\Lambda_{1}}{2}}e^{-\sqrt{\frac{\beta_{d}}{\bar{\gamma_{d}}\zeta_{d}^{2}}\gamma}},
\end {align}
where $\beta_{d}$ defines the path loss of the RIS-aided path and $\bar{\gamma_{d}}$ denotes the average SNR of $\mathcal{S}-\mathcal{R}_{k}-\mathcal{D}$ link.

%%%%%%%%%%%%%%%% END OF EQUATION %%%%%%%%%%%%%%
%%%%%%%%%%%% PROOF OF RIS PDF %%%%%%%%%%%%%
\textit{Proof}: Let $Y_{k}=\theta_{k}\phi_{k}$ be the product of two $\alpha$-$\mu$ random variables (RVs) where $\theta_{k}$ denotes the $\mathcal{S}$-$\mathcal{R}_{k}$ hop with $\alpha_{s}$ and $\beta_{s}$ being the fading parameters, and $\Omega_{s}$ be the scale parameter whereas $\phi_{k}$ defines the $\mathcal{R}_{k}$-$\mathcal{D}$ hop with $\alpha_{r}$ and $\beta_{r}$ being the fading parameters, and $\Omega_{r}$ be the scale parameter. Therefore, the PDF of $Y_{k}$ can be expressed as
%%%%%%%%%%%%%%%%%% Fading PDF Expression %%%%%%%%%%%%
\begin{align}\label{eqn:3}
f_{Y_{k}}(y)=\chi_{1}y^{\chi_4}K_{\chi_{2}}(\chi_{5}y^{\frac{\alpha_{s}}{2}}), 
\end{align}
%%%%%%%%%%%%%%%%%%% END Expression %%%%%%%%%%%%%%%%%
where $\chi_{1}=\frac{2\alpha_{s}\alpha_{r}\mu_{s}^{\mu{s}}\mu_{r}^{\mu{r}}}{\alpha_s \Gamma{(\mu_{s})}\Gamma{(\mu_{r})}\Omega_{s}^{\alpha_{s}\mu_{s}}\Omega_{r}^{\alpha_{r}\mu_{r}}}(\frac{\mu_{r}\Omega{s}^{\alpha_{s}}}{\mu_{s}\Omega{r}^{\alpha_{r}}})^{\frac{\alpha_{s}\mu_{s}-\alpha_{r}\mu_{r}}{2\alpha}}, \chi_{2}=\frac{\alpha_{s}\mu_{s}-\alpha_{r}\mu_{r}}{\alpha}$, $\chi_{3}=\frac{\mu_{r}\mu_{s}}{\Omega{r}^{\alpha_{r}}\Omega{s}^{\alpha_{s}}}$, $\chi_{4}=\frac{\alpha_{s}\mu_{s}+\alpha_{r}\mu_{r}}{2}-1$, $\chi_{5}=2\sqrt{\chi_{3}}$, and $K_{\nu}(\cdot)$ denotes the modified $\nu^{th}$-order Bessel function \cite[Eq. (8.432)]{gradshteyn2014table}. However, with the help of \cite[Sec.~(2.2.2)]{primak2005stochastic}, it can be stated that the PDF of $\Tilde{Y}=\sum_{k=1}^{N_{d}}Y_{k}$ is tightly approximated by the first term of a Laguerre expansion. Hence, the corresponding PDF can be expressed as
%%%%%%%%%%%%%%%%%%%% PDF Expression %%%%%%%%%%%%%%
\begin{align}\label{eqn:4}
f_{\Tilde{Y}}(y)\simeq\frac{1}{\Gamma{(\vartheta_{d}})\zeta_{d}^{\vartheta_{d}}}y^{\vartheta_{d}-1}e^{-\frac{y}{\zeta_{d}}}, 
\end{align}
%%%%%%%%%%%%%%%%%%%% END Expression %%%%%%%%%%%%%%%
where $\vartheta_{d}$ denotes the shape parameter and $\zeta_{d}$ defines the scale parameter, respectively, and they are related to the $s^{th}$ moment of $Y_{k}$, can be expressed as
%%%%%%%%%%% kth moment expression %%%%%%%%%%%%%
\begin {align}\label{eqn:5}
\mathbb{E}(Y_{k}^{s})=\chi_{1}\frac{2^{s+\chi_{4}-1}}{\chi_{5}^{(s+\chi_{4}+1)}}\Gamma(\frac{\chi_{6}+s}{2})\Gamma(\frac{\chi_{7}+s}{2}), 
\end {align}
%%%%%%%%%%%%%%%%%%%% End %%%%%%%%%%%%%%%%%%%%
where $\chi_{6}=1+\chi_{4}+\chi_{2}$, $\chi_{7}=1+\chi_{4}-\chi_{2}$, and $\Gamma(\cdot)$ is the Gamma function as defined in \cite[Eq. (8.310)]{gradshteyn2014table}. Now, utilizing the RVs transformation technique and performing some mathematical manipulations, the PDF of $\gamma_{d}$ can be finally shown as \eqref{eqn:1}. Therefore, the CDF of $\gamma_{d}$ can be written mathematically as
%%%%%%%%%%%%%%%%%% CDF Expression %%%%%%%%%%%%%%
\begin{align}\label{eqn:1.1}
F_{\gamma_{d}}(\gamma)=\int_{0}^{\gamma}f_{\Tilde{Y}}(y)dy.
\end{align}
%%%%%%%%%%%%%%%%%%% End Expression %%%%%%%%%%%%%%
Substituting \eqref{eqn:1} into \eqref{eqn:1.1} and utilizing \cite[Eq. (3.381.8)]{gradshteyn2014table}, the CDF can be obtained. Thus, the proof is completed.
%%%%%%%%%%%%%%%%%%%%%%%%%%%%%%%%%%%%%%%%%%%%%

%%%%%%%%%%%%%%%%%%%%%%%%%%%%%%%%%%%%%%%%%%%%%
\subsection{PDF and CDF of $\gamma_{d,i}$}
Assuming an interference limited system, the PDF of $\gamma_{d,i}$ is expressed as
%%%%%%%%%%%%%%% RIS + CCI PDF %%%%%%%%%%%%%%%
\begin {align}\label{eqn:6}
f_{\gamma_{d, i}}(\gamma)=\Xi_2\gamma^{\frac{\vartheta_d}{2}-1}G_{\mathcal{B}_1, 2}^{2, \mathcal{B}_1}\left[\mathcal{A}_1\gamma \bigg|
\begin{array}{c}
\alpha_i-\omega_0-1\\
0, \frac{1}{2}\\
\end {array}
\right], 
\end {align}
%%%%%%%%%%%%%%%%%%%% END EQUATION %%%%%%%%%%%%%%%%%
where $\mathcal{A}_{1}$=$\frac{\Xi_1^{2}(\mathcal{B}_1 \bar{\gamma_i}^{\alpha_i})^{\mathcal{B}_1}}{4(\mathcal{M}\mu_i)^{\mathcal{B}_1}}$, $\mathcal{B}_1=\frac{1}{\alpha_i}$,  $\omega_0=\frac{\vartheta_d}{2}+\mathcal{M}\alpha_i\mu_i-1$, $\Xi_2=\frac{(\mathcal{M}\mu_i)^{\mathcal{M}\mu_i-\mathcal{B}_1(\omega_0+1)}\mathcal{B}_1^{\mathcal{B}_1(\omega_0+1)+\frac{1}{2}}\bar{\gamma}_i^{\alpha_i\mathcal{B}_1(\omega_0+1)-\mathcal{M}\alpha_i\mu_i}\alpha_i\Xi_1}{\sqrt{\pi}\sqrt{2}^{\mathcal{B}_1-1}\Gamma{(\mathcal{M}\mu_i)}}$, $\alpha_i$ and $\mu_{i}$ are the fading parameters of $\mathcal{M}_{i}-\mathcal{D}$ link, $\bar{\gamma_{i}}$ denotes the average SNR of $\mathcal{M}_{i}-\mathcal{D}$ link and $G[\cdot]$ is the Meijer's G function defined in \cite{badrudduza2021security}. Accordingly, the CDF of $\gamma_{d, i}$ is expressed as
%%%%%%%%%%%%%%% RIS + CCI LIMITED CDF %%%%%%%%%%%%%%%
\begin {align}\label{eqn:7}
F_{\gamma_{d, i}}(\gamma)=\Xi_2 \gamma^{\frac{\vartheta_d}{2}}G_{\mathcal{B}_1+1, 3}^{2, \mathcal{B}_1+1}\left[\mathcal{A}_1\gamma \bigg|
\begin {array}{c}
\frac{-\vartheta_d}{2}+1, -\omega_0\\
0, \frac{1}{2}, \frac{-\vartheta_d}{2}
\end {array}
\right].
\end {align}
%%%%%%%%%%%%%%%%%%% END EQUATION %%%%%%%%%%%%%%%%%%%%%
\textit{Proof:} Assuming all the interferers with equal power case, the PDF of $\mathcal{M}_{i}$-$\mathcal{D}$ link can be expressed as
%%%%%%%%%%%%%%%%%% CCI PDF %%%%%%%%%%%%%%%%
\begin {align}\label{eqn:8}
f_{\gamma_{i}}(x)=\frac{\alpha_{i}(\mathcal{M}\mu_i)^{\mathcal{M}\mu_i}}{\Gamma(\mathcal{M}\mu_i)\bar{\gamma}_i^{\alpha_{i}\mathcal{M}\mu_{i}}}x^{\alpha_{i}\mathcal{M}\mu_{i-1}}e^{-\frac{\mathcal{M}\mu_{i}}{\bar{\gamma}_i^{\alpha_i}}x^{\alpha_{i}}}.
\end {align}
%%%%%%%%%%%%%%%%%%%%%% END EQUATION %%%%%%%%%%%%%%
Now, the PDF of $\gamma_{d, i}$ can be written as 
\begin {align}\label{eqn:9}
%%%%%%%%%%%%%%%%%%%% CCI FORMULA %%%%%%%%%%%%%%%%%
f_{\gamma_{d,i}}(\gamma)=\int_{0}^{\infty}x f_{\gamma_{d}}(\gamma x)f_{\gamma_{i}}(x)dx.
\end {align}
%%%%%%%%%%%%%%%%%%%%% END EQUATION %%%%%%%%%%%%%%%%
Substituting \eqref{eqn:1} and \eqref{eqn:8} into \eqref{eqn:9}, the PDF of $\gamma_{d,i}$ as in \eqref {eqn:6} is obtained. Subsequently, replacing \eqref{eqn:6} into \eqref{eqn:1.1} and with the aid of \cite[Eq. (2.24.2.2)]{prudnikov1988integrals}, $F_{\gamma_{d,i}}$ can be derived. This concludes the proof.
%%%%%%%%%%%%%%%%%%%%%%%%%%%%%%%%%%%%%%%%%%%%%%%%%%
\subsection{PDF of $\gamma_{e}$}
\textit{Scenerio-I:} Here, we consider all the $\mathcal{S}$-$\mathcal{E}$ link undergo $\alpha$-$\mu$ fading distributions. Based on the eavesdropping characteristic, colluding (\textit{Case-I}) and non-colluding (\textit{Case-II}) cases are considered.
\\
\textit{Case-I:} In the colluding eavesdropper scheme, eavesdroppers collaborate in order to intercept legitimate information. Hence, the PDF of $\gamma_{e,c}$ can be expressed as
%%%%%%%%%%%%%% Colluding PDF Without RIS %%%%%%%%
\begin{align}\label{eqn:10}
f_{\gamma_{e, c}}(\gamma)=\Xi_3\gamma^{(\mathcal{L}\alpha_e\mu_e-1)}e^{-\Xi_4\gamma^{\alpha_e}}, 
\end{align}
%%%%%%%%%%%%%%% END EQUATION %%%%%%%%%%%%%%%%%%
where $\alpha_e$ and $\mu_e$ denote the fading parameters of $\mathcal{S}-\mathcal{E}_{q}$ links, $\Xi_3=\frac{\alpha_e(\mathcal{L}\mu_e)^{\mathcal{L}\mu_{e}}}{\Gamma(\mathcal{L}\mu_{e})\bar{\gamma_{e}}^{\alpha_{e}\mathcal{L}\mu_{e}}}$, $\Xi_4=\frac{\mathcal{L}\mu_{e}}{\bar{\gamma_{e}}^{\alpha_{e}}}$, and $ \bar{\gamma_e}$ denotes the average SNR of each links.

\textit{Case-II:}
Typically, in a NC situation, each eavesdropper performs an individual decoding process to salvage the legitimate signal. Therefore, the PDF of $\gamma_{e,nc}$ can be expressed as
%%%%%%%%%%%%% NC PDF Without RIS %%%%%%%%%%
\begin{align}\label{eqn:12}\nonumber
f_{\gamma_{e, nc}}(\gamma)&=\sum_{\Lambda_3=0}^{\infty}\sum_{k_1+...+k_{\mu_e-1}=\Lambda_3}\Xi_5\gamma^{\frac{\alpha_e\mu_e}{2}-1+\Xi_6}
\\ \times
&e^{-(\Lambda_3+1)\mu_e(\frac{\gamma}{\bar{\gamma_e}})^\frac{\alpha_e}{2}}, 
\end{align}
%%%%%%%%%%%%%%%%% END EQUATION %%%%%%%%%%%%%%%%
where $\Xi_5=\binom{\mathcal{L}-1}{\Lambda_3}\binom{\Lambda_3}{k_1...k_{\mu_e-1}}\frac{\alpha_e\mu_e^{\mu_e+\Lambda_2\Lambda_3}\mathcal{L}}{2\Gamma{\mu_e}(\Lambda_2!)^{\Lambda_3}(-1)^{\Lambda_3}}\bar{\gamma_e}^{-\frac{\alpha_e\Lambda_2\Lambda_3+\alpha_e\mu_e}{2}}$ and $\Xi_6=\sum_{\Lambda_2}\frac{\alpha_e}{2}\Lambda_2k_{\Lambda_2}$.
\\
\textit{Proof:}
The PDF for NC eavesdropper can be written mathematically as
%%%%%%%%%%%%%%% FORMULA OF NC EQUATION %%%%%%%%%%%
\begin{align}\label{eqn:12.1}
f_{\gamma_{e, nc}}(\gamma)=\mathcal{L}\left[F_{\gamma_{e, c}}(\gamma)\right]^{\mathcal{L}-1}f_{\gamma_{e, c}}(\gamma).
\end{align}
%%%%%%%%%%%%%%%%% END EQUATION %%%%%%%%%%%%%%%%%
%With the help of [\cite{lei2017secrecy}, eq.(1)] and [......] the \eqref{eqn:12.1} will
%\begin{align}\nonumber
%=& L\left[1-\sum_{\Lambda_2=0}^{\mu_e-1}\frac{\left(\mu_e\bar{\gamma_e}^{\frac{-\alpha_e}{2}}\right)^{\Lambda_2}}{\Lambda_2!}\gamma^{\frac{\alpha_e \Lambda_2}{2}} e^{-\mu_e(\frac{\gamma}{\bar{\gamma_e}})^\frac{\alpha_e}{2}}\right]^{L-1}
%\\ \times
% &\frac{\alpha_e\mu_e^{\mu_e}\gamma^{\frac{\alpha_e\mu_e}{2}-1}e^{-\mu_e(\frac{\gamma}{\bar{\gamma_e}})^\frac{\alpha_e}{2}}}{2\Gamma{\mu_e\bar{\gamma}_e^{\frac{\alpha_e\mu_e}{2}}}}.
% \end{align}
With the help of \cite [Eqs. (4-5)]{juel2021secrecy} and \cite[Eq. (1.110)]{gradshteyn2014table}, and then performing some algebraic manipulations, we obtain
\begin{align}\label{eqn:12.2}\nonumber
f_{e,nc}=&\sum_{\Lambda_3=0}^{\infty}\binom{\mathcal{L}-1}{\Lambda_3}\frac{\mathcal{L}\alpha_e\mu_e^{\mu_e+\Lambda_2\Lambda_3}}{2\Gamma{\mu_e}(\Lambda_2!)^{\Lambda_3}(-1)^{\Lambda_3}}\bar{\gamma_e}^{-\frac{\alpha_e\Lambda_2\Lambda_3+\alpha_e\mu_e}{2}}
\\ \times
&\gamma^{\frac{\alpha_e\mu_e}{2}-1}e^{-(\Lambda_3+1)\mu_e(\frac{\gamma}{\bar{\gamma_e}})^\frac{\alpha_e}{2}}\left(\sum_{\Lambda_2=0}^{\mu_e-1}\gamma^{\frac{\alpha_e\Lambda_2}{2}}\right)^{\Lambda_3}.
\end{align}
Now, the expansion of $\left(\sum_{\Lambda_2=0}^{\mu_e-1}\gamma^{\frac{\alpha_e\Lambda_2}{2}}\right)^{\Lambda_3}$ is written as
\begin{align}\label{eqn:12.3}\nonumber
\left(\sum_{\Lambda_2=0}^{\mu_e-1}\gamma^{\frac{\alpha_e\Lambda_2}{2}}\right)^{\Lambda_3}=&\sum_{k_1+..+k_{\mu_e-1}=\Lambda_3}\binom{\Lambda_3}{k_1, ..., k_{\mu_e-1}}
\\ \times
&\gamma^{\frac{\alpha_e}{2}\sum_{\Lambda_2}\Lambda_2k_{\Lambda_2}}.
\end{align}
Finally, substituting \eqref{eqn:12.3} into \eqref{eqn:12.2}, the proof is completed.

\textit{Scenario-II:}
Here, through $\mathcal{I}_{p}$, the eavesdroppers gain access to the information they need to overhear. It is important to note that $\mathcal{D}$ and $\mathcal{E}$ use two different RISs, i.e., $\mathcal{R}_{k}$ and $\mathcal{I}_{p}$, respectively.

\textit{Case-I:}
According to \eqref{eqn:1} and \eqref{eqn:2}, the PDF and CDF for colluding eavesdroppers can be respectively expressed as
%%%%%%%%%%%%%%%%% PDF & CDF OF COLLUDING EAV WITH RIS %%%%%%%%%%
\begin{align}\label{eqn:13}
f_{\gamma_{E, C}}(\gamma)&=\Xi_7\gamma^{\frac{\vartheta_E}{2}-1}e^{-\Xi_8\sqrt{\gamma}}, 
\\\label{eqn:13.1}
F_{\gamma_{E, C}}(\gamma)&=1-\sum_{\Lambda_4=0}^{\vartheta_E-1}\frac{\Xi_8^{\Lambda_4}}{\Lambda_4!}\gamma^{\frac{\Lambda_4}{2}}e^{-\Xi_8\sqrt{\gamma}}, 
\end{align}
%%%%%%%%%%%%%%%% END EQUATION %%%%%%%%%%%%%
where $\Xi_7=(\frac{\beta_{E}}{\bar{\gamma}_E})^{\frac{\vartheta_{E}}{2}}\frac{\zeta_{E}^{-\vartheta_{E}}}{2\Gamma{(\vartheta_{E})}}$, $\Xi_8=\sqrt{\frac{\beta_{E}}{\bar{\gamma}_E}\zeta_{E}^{2}}$, $\beta_{E}$ and $\vartheta_{E}$ are the shape and scale parameter of $\mathcal{S}-\mathcal{I}_{p}-\mathcal{E}_{q}$ links, respectively and $\bar{\gamma}_E$ is the average SNR.
 
 \textit{Case-II:}
 With the help of a similar process followed while deriving \eqref{eqn:12}, the PDF for NC eavesdropper can be expressed as
 %%%%%%%%%%%%%%%%%% PDF OF NC EAV %%%%%%%%%%%%%%
 \begin{align}\label{eqn:13.2}\nonumber
f_{\gamma_{E, NC}}(\gamma)=&\sum_{\Lambda_5=0}^{\infty}\sum_{r_1+...+r_{(\vartheta_E-1)}=\Lambda_5}\Xi_{10}\gamma^{\frac{\vartheta_E}{2}-1+\Xi_9}
\\ \times
&e^{-(\Xi_8+\Lambda_5\Xi_8)\sqrt{\gamma}}, 
 \end{align}
 %%%%%%%%%%%%%%%%%%% END EQUATION %%%%%%%%%%%%
 where $\Xi_{9}=\sum_{\Lambda_{4}}\frac{\Lambda_{4} r_{\Lambda_{4}}}{2}$ and $\Xi_{10}=\binom{L-1}{\Lambda_{5}}\binom{\Lambda_{5}}{r_1...r_{\vartheta_{E-1}}}(\frac{\Xi_{8}^{\Lambda_{4}}}{\Lambda_{4}!})^{\Lambda_{5}}(-1)^{\Lambda_{5}}\mathcal{L}\Xi_{7}$. Note that assuming $\beta_{E}=\beta_{d}$, $\vartheta_{E}=\vartheta_{d}$, $\zeta_{E}=\zeta_{d}$, $\bar{\gamma}_{E}=\bar{\gamma}_{d}$, and $\mathcal{N}_{E}=\mathcal{N}_{d}$ the $\textit{Scenerio-II}$ reduces to $\textit{Scenerio-III}$.

 %%%%%%%%%%%%%%%%%%%%% 1st one %%%%%%%%%%%%
\setcounter{eqnback}{\value{equation}} \setcounter{equation}{24}
\begin{figure*}[!t]
\vspace{0mm}
\begin{align}\label{eqn:16.1.1}
SOP_{\gamma_{e, c}}^{(\infty)}=\sum_{\kappa_1=1}^{2\alpha_e}\Xi_2\Xi_3\mathcal{A}_1^{\omega_1-1}\mathcal{A}_2^{\mathcal{S}_{1,\kappa_1}-1}\alpha_{e}^{\omega_0-\omega_1-1}\Psi^{-\alpha_{e}L\mu_{e}} 
\frac{\prod_{p_1=1;p_1\neq\kappa_1}^{2\alpha_e}\Gamma(\mathcal{S}_{1,\kappa_1}-\mathcal{S}_{1,p_1})\prod_{p_1=1}^{1+2\alpha_e}\Gamma(1+\mathcal{S}_{2,p_1}-\mathcal{S}_{1,\kappa_1})}{\prod_{p_1=1+2\alpha_e}^{3\alpha_e}\Gamma(1+\mathcal{S}_{1,p_1}-\mathcal{S}_{1,\kappa_1})}.
\end{align}
%\hrulefill
\end{figure*}
%%%%%%%%%%%%%%%%% 2nd %%%%%%%%%%%%%%%%%%%%%
%\vspace{-5mm}
\begin{figure*}[!t]
\vspace{-7mm}
\begin{align}\label{eqn:16.1.2}
SOP_{\gamma_{e, nc}}^{(\infty)}&=\sum_{k_1+...+k_{\mu_e-1}=\Lambda_3}\sum_{\Lambda_3=0}^{\infty}\sum_{\kappa_2=1}^{2}
%\\ \times \nonumber
 \frac{2\Xi_5\Xi_2\sqrt{B_2\pi}\prod_{p_2=1;p_2\neq\kappa_2}^{2}\Gamma(\mathcal{S}_{3,\kappa_2}-\mathcal{S}_{3,p_2})\prod_{p_2=1}^{2\mathcal{B}_2+\mathcal{B}_1+1}\Gamma(1+\mathcal{S}_{4,p_2}-\mathcal{S}_{3,\kappa_2})}{(2\pi)^{B_2}\Psi^{(\frac{\alpha_e\mu_e}{2}+\Xi_6)}\mathcal{A}_1^{(\frac{\alpha_e\mu_e+\vartheta_d}{2}+\Xi_6)}\mathcal{A}_3^{-\mathcal{S}_{3,\kappa_2}+1}\Gamma(1+\mathcal{S}_{3,3}-\mathcal{S}_{3,\kappa_2})}.
\end{align}
%\hrulefill
\end{figure*}
%%%%%%%%%%%%%%%%%%%% 3rd %%%%%%%%%%%%%%%%
\begin{figure*}[!t]
\vspace{-7mm}
\begin{align}\label{eqn:17.1}
&SOP_{\gamma_{E, C}}^{(\infty)}=\sum_{\kappa_3=1}^{2}\Xi_2\Xi_7\mathcal{A}_1^{\omega_2-1}\mathcal{A}_4^{\mathcal{S}_{5,\kappa_3}-1}\pi^{-\frac{1}{2}}\Psi^{-\frac{\vartheta_E}{2}}
%\\ \times
%&
\frac{\prod_{p_3=1;p_3\neq\kappa_3}^{2}\Gamma(\mathcal{S}_{5,\kappa_3}-\mathcal{S}_{5,p_3})\prod_{p_3=1}^{\mathcal{B}_1+3}\Gamma(1+\mathcal{S}_{6,p_3}-\mathcal{S}_{5,\kappa_3})}{\Gamma(1+\mathcal{S}_{5,3}-\mathcal{S}_{5,\kappa_3})}.
\end{align}
%\hrulefill
\end{figure*}
%%%%%%%%%%%%%%%%%%%%%%%%% 4th %%%%%%%%%%%%
\begin{figure*}[!t]
\vspace{-8mm}
\begin{align}
\label{eqn:18.1}
SOP_{\gamma_{E, NC}}^{(\infty)}=\sum_{\Lambda_5=0}^{\infty}\sum_{r_1+...+r_{(\vartheta_E-1)=\Lambda_5}}\sum_{\kappa_4=1}^{2}
\frac{\Xi_{10}\Xi_2\prod_{p_4=1;p_4\neq\kappa_4}^{2}\Gamma(\mathcal{S}_{7,\kappa_4}-\mathcal{S}_{7,p_4})\prod_{p_4=1}^{\mathcal{B}_1+3}\Gamma(1+\mathcal{S}_{8,p_4}-\mathcal{S}_{7,\kappa_4})}{\sqrt{\pi}\Psi^{\frac{\vartheta_E}{2}+\Xi_9}\mathcal{A}_1^{-\omega_2+\Xi_9+1}\mathcal{A}_5^{-\mathcal{S}_{7,\kappa_4}+1}\Gamma(1+\mathcal{S}_{7,3}-\mathcal{S}_{7,\kappa_4})}.
\end{align}
\hrulefill
\vspace{-6mm}
\end{figure*}

%%%%%%%%%%%%%%%%<SECTION>%%%%%%%%%%%%%%%
\section{Performance Analysis}

\label{metrics}
In this section, the novel analytical expressions of SOP are demonstrated by considering colluding and NC eavesdropping attacks. 
\subsection{SOP Analysis}
To achieve secure communication, the instantaneous secrecy capacity ($C_s$), must be greater than the predetermined target secrecy rate ($\tau_{0}$), otherwise, an SOP event occurs. Mathematically, the lower bound of SOP can be written as
%%%%%%%%%%%%%%%%%% SOP FORMULA %%%%%%%%%%%
\setcounter{eqnback}{\value{equation}} \setcounter{equation}{18}
\begin {align}\label{eqn:15}
SOP\geq SOP_L=\int_{0}^{\infty}F(\Psi\gamma)f(\gamma)d\gamma.
\end{align}
%%%%%%%%%%%%%%%%%% END EQUATION %%%%%%%%%%%%
where $\Psi=2^{\tau_{0}}$.
\\
\textit{Scenario-I:Case-I:} Substituting \eqref{eqn:7} and \eqref{eqn:10} into \eqref{eqn:15}, we obtain
\begin{align}\nonumber
SOP_{\gamma_{e, c}}=&\Xi_2\Xi_3\Psi^{\frac{\vartheta_d}{2}}\int_{0}^{\infty}\gamma ^{-\omega_1}G_{0, 1}^{1, 0}\left[\Xi_4\gamma^{a_{e}}\bigg|
\begin{array}{c}
-\\
0
\end{array}
\right]
\\ \times 
&G_{\mathcal{B}_1+1, 3}^{2, \mathcal{B}_1+1}\left[\mathcal{A}_1\gamma\Psi \bigg|
\begin {array}{c}
-\frac{\vartheta_d}{2}+1, -\omega_0\\
0, \frac{1}{2}, -\frac{\vartheta_d}{2}
\end {array}
\right] d\gamma.
\end{align}
Utilizing the identity \cite[Eq. (2.24.1.1)]{prudnikov1988integrals} and performing some mathematical manipulations, the lower bound of SOP is expressed as
%%%%%%%%%%%%%%%%%% SOP EXPRESSION FOR COLLUDING EAV WITHOUT RIS %%%%%%%%%%%%%%%%%
\begin{align}\label{eqn:16.1}
SOP_{\gamma_{e, c}}=&\frac{\Xi_2\Xi_3\mathcal{A}_1^{\omega_1-1}}{\alpha_{e}^{1-\omega_0+\omega_1}\Psi^{\alpha_{e}\mathcal{L}\mu_{e}}} 
%\\ \times
G_{3\alpha_{e}, 1+2\alpha_{e}}^{1+2\alpha_{e}, 2\alpha_{e}}\left[\mathcal{A}_2\bigg|
\begin{array}{c}
\mathcal{S}_1\\
\mathcal{S}_2
\end{array}
\right],  
\end{align}
%%%%%%%%%%%%%%%%%%%%% END EQUATION %%%%%%%%%%%%
where $\mathcal{A}_2=\frac{\Xi_4\alpha_{e}^{\alpha_{e}}}{(\mathcal{A}_1\Psi)^{\alpha_{e}}}$, $\omega_1=-\frac{\vartheta_d}{2}-\mathcal{L}\alpha_e\mu_e+1$, $t=-L\alpha_e\mu_e+1$, $\mathcal{S}_1=\triangle(\alpha_e, \omega_1), \triangle(\alpha_e, \omega_1-\frac{1}{2}), \triangle(\alpha_e, t)$, and $\mathcal{S}_2=0, \triangle(\alpha_e, -\mathcal{L}\alpha_e\mu_e), \triangle(\alpha_e, \omega_1+\omega_0)$.

\textit{Case-II:} By plugging \eqref{eqn:7} and \eqref{eqn:12} into \eqref{eqn:15} and utilizing the identity of \cite[Eq. (2.24.1.1)]{prudnikov1988integrals}, the SOP for NC eavesdroppers can be obtained as
%%%%%%%%%%%%%%% SOP EXPRESSION FOR NC EAV WITHOUT RIS %%%%%%%%%%%%%%%
\begin{align}\label{eqn:16}\nonumber
&SOP_{\gamma_{e, nc}}=\sum_{k_1+...+k_{\mu_e-1}=\Lambda_3}\sum_{\Lambda_3=0}^{\infty}2\Xi_5\Xi_2\sqrt{\mathcal{B}_2\pi}(2\pi)^{-\mathcal{B}_2}
\\ &\times 
\frac{\Psi^{-(\frac{\alpha_e\mu_e}{2}+\Xi_6)}}{\mathcal{A}_1^{(\frac{\alpha_e\mu_e+\vartheta_d}{2}+\Xi_6)}}
%\\ \times
%& 
G_{3, 2\mathcal{B}_2+\mathcal{B}_1+1}^{2\mathcal{B}_2+\mathcal{B}_1+1, 2}\left[\mathcal{A}_3 \bigg|
\begin{array}{c}
\mathcal{S}_3\\
\mathcal{S}_4
\end{array}
\right], 
\end{align}
%%%%%%%%%%%%% END EQUATION %%%%%%%%%%%%%
where $\mathcal{A}_3=\frac{[(1+\Lambda_3)\mu_e\bar{\gamma}^{-\frac{\alpha_e}{2}}]^{2\mathcal{B}_2}}{(2\mathcal{B}_2)^{2\mathcal{B}_2}\mathcal{A}_1\Psi}, \mathcal{B}_2=\frac{1}{\alpha_e},\mathcal{S}_3=\triangle(1, -\frac{\vartheta_d}{2}-\frac{\alpha_e\mu_e}{2}+1-\Xi_6)$, and $\mathcal{S}_4=\triangle(2\mathcal{B}_2, 0), \triangle(1, -\frac{\alpha_e\mu_e}{2}-\Xi_6)$.
\\
\textit{Scenerio-II:Case-I:} Substituting \eqref{eqn:7} and \eqref{eqn:13} into \eqref{eqn:15} and following the similar process of \eqref{eqn:16}, we obtain
%%%%%%%%%%%%% SOP EXPRESIONS FOR COLLUDING EAV WITH RIS %%%%%%%%%%%%%%%%
\begin{align}\label{eqn:17}
SOP_{\gamma_{E, C}}=&\frac{\Xi_2\Xi_7\mathcal{A}_1^{\omega_2-1}}{\pi^{\frac{1}{2}}\Psi^{\frac{\vartheta_E}{2}}}
%\\ \times
%&
G_{3, \mathcal{B}_1+3}^{\mathcal{B}_1+3, 2}\left[\mathcal{A}_4\bigg|
\begin{array}{c}
\mathcal{S}_5\\
\mathcal{S}_6
\end{array}
\right], 
\end{align}
%%%%%%%%%%%%%%%%%%%% END EQUATION %%%%%%%%%%%%%%
where $\mathcal{A}_4=\frac{\Xi_8^{2}}{4\mathcal{A}_1\Psi}$, $\omega_2=-\frac{\vartheta_d}{2}-\frac{\vartheta_E}{2}+1$, $\mathcal{S}_5=\triangle(1, \omega_2)$, and $\mathcal{S}_6=\triangle(2, 0), \triangle(1, -\frac{\vartheta_E}{2})$.
\\
\textit{Case-II:} Similar to \eqref{eqn:17}, replacing \eqref{eqn:7} and \eqref{eqn:13.2} into \eqref{eqn:15}, the lower bound of SOP for NC eavesdroppers with RIS can be obtained as
%%%%%%%%%%%%%%%%% SOP EXPRESSION FOR NC EAV WITH RIS %%%%%%%%%%%%%%%%
\begin{align}\label{eqn:18}\nonumber
SOP_{\gamma_{E, NC}}=&\sum_{\Lambda_5=0}^{\infty}\sum_{r_1+...+r_{(\vartheta_E-1)=\Lambda_5}}\frac{\Xi_{10}\Xi_2}{\sqrt{\pi}\Psi^{\frac{\vartheta_E}{2}+\Xi_9}}\mathcal{A}_1^{\omega_2-\Xi_9-1}
\\ \times
& G_{3, \mathcal{B}_1+3}^{\mathcal{B}_1+3, 2}\left[\mathcal{A}_5\bigg|
\begin{array}{c}
\mathcal{S}_7\\
\mathcal{S}_8
\end{array}
\right], 
\end{align}
%%%%%%%%%%%%%%%% END EQUATION %%%%%%%%%%%%%
where $\mathcal{A}_5=\frac{(\Xi_8+\Lambda_5\Xi_8)^2}{4\Psi\mathcal{A}_1}$, $\mathcal{S}_7=\triangle(1, \omega_2-\Xi_9)$, and $\mathcal{S}_8=\triangle(2, 0), \triangle(1, -\frac{\vartheta_E}{2}-\Xi_9)$.

%%%%%%%%%%%%%%%%%%%%
\subsection{Asymptotic Analysis}
The exact SOP in section A does not provide much insight
into the secure outage performance of the proposed model.
Hence, we also derive the asymptotic expression of SOP
in this section. Herein, Meijer’s $G$ term is converted into
an asymptotic term  by utilizing \cite[Eq. (41)]{ansari2015performance} as expressed in \eqref{eqn:16.1.1}, \eqref{eqn:16.1.2}, \eqref{eqn:17.1}, and \eqref{eqn:18.1}.

%*************<END-EQUATION>************

%%%%%%%%%%%%%%%%<SECTION>%%%%%%%%%%%%%%%
\section{Numerical Results}\label{results}

\begin{figure}[b!]
\vspace{0mm}
\centerline{\includegraphics[width=0.4\textwidth]{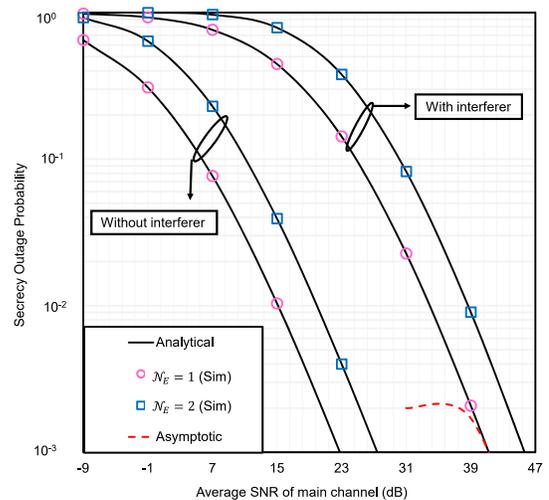}}
    \vspace{-2mm}
    \caption{SOP vs $\bar{\gamma}_d$ for the selected values of $\mathcal{N}_E$ with $\mathcal{M}=20$.}
  \label{r1}  
  \vspace{-2mm}
\end{figure}
In this section, the effects of the system parameters are presented through graphs generated using analytical expressions. The validity of each graph is confirmed through 100,000 random channel sample simulations performed in MATLAB. The system parameters are set to $\alpha_s=\alpha_r=2, \alpha_i=\alpha_e=1$, $\mu_s=\mu_r=\mu_i=\mu_e=1$, $\mathcal{N}_d=\mathcal{N}_{E}=2$, $\mathcal{L}=2$, $\mathcal{M}=2$, $\beta_d=\beta_E=0.5$, $\tau_0=0.1$,$\Omega_s=\Omega_r=1$,$\bar{\gamma}_i=100$ dB, and $\bar{\gamma}_e=\bar{\gamma}_E=1$ dB, unless specified otherwise. Additionally, asymptotic analysis is conducted showing a tight match at a high SNR regime with the analytical one. Furthermore, the correctness of the derived expressions is confirmed by close matches between analytical results and simulation results. Note that in all figures, "Sim" indicates the simulation results.

Figure \ref{r1} illustrates the effects of interference and $\mathcal{N}_{E}$ on secure outage performance. As is evident, the security performance worsens with interference and an increase in the number of reflecting elements in $\mathcal{I}$. This occurs because interference reduces the SINR at $\mathcal{D}$ and the eavesdropper link becomes stronger as $\mathcal{N}_{E}$ increases.

%%%%%%%%%%%%%%%%%%%%%%%%%%%%%%%%%%%%%%%%%%%%
\begin{figure}[h!]
\vspace{0mm}
\centerline{\includegraphics[width=0.4\textwidth]{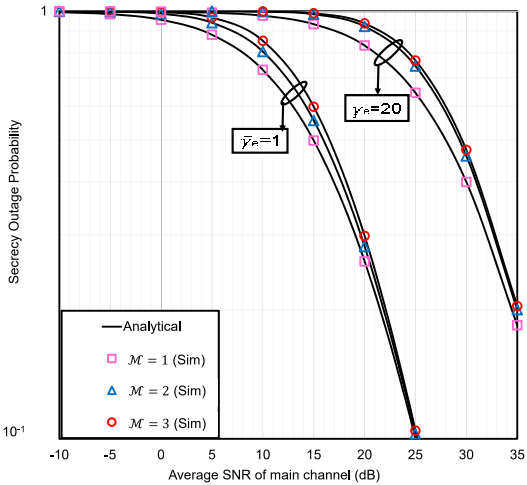}}
    \vspace{-3mm}
    \caption{SOP vs $\bar{\gamma}_d$ for the selected values of $\mathcal{M}$ and $\bar{\gamma}_e$. }
  \label{r2}
  \vspace{-3mm}
\end{figure}
The illustration in Fig. \ref{r2} demonstrates the impact of the number of interferers at $\mathcal{D}$.
It is clear that as the number of $\mathcal{M}$ increases, the SOP declines due to the increase in interference that lowers the SINR at $\mathcal{D}$. Additionally, a rise in $\bar{\gamma}_e$ strengthens the eavesdropper link, resulting in a higher secrecy outage.

%%%%%%%%%%%%%%%%%%%%%%%%%%%%%%%%%%%%%%%%%%%%
\begin{figure}[!h]
\vspace{0mm}
\centerline{\includegraphics[width=0.4\textwidth]{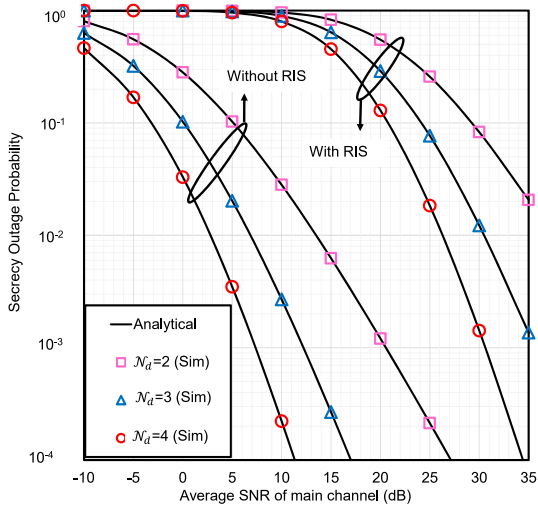}}
    \vspace{0mm}
    \caption{SOP vs $\bar{\gamma}_d$ for the selected values of $\mathcal{N}_d$ with $\mathcal{N}_E=20$.}
  \label{r3}  
  \vspace{0mm}
\end{figure}
Figure \ref{r3} demonstrates the significance of RIS and reflecting elements in $\mathcal{R}$ showing that a degraded secrecy performance is achieved when RIS is present in the eavesdropper link as opposed to a direct link without RIS. Additionally, as the number of $\mathcal{N}_{d}$ increases, there is an improvement in the SNR at $\mathcal{D}$, accompanied by a reduction in the SOP.

\begin{figure}[!h]
\vspace{0mm}
\centerline{\includegraphics[width=0.4\textwidth]{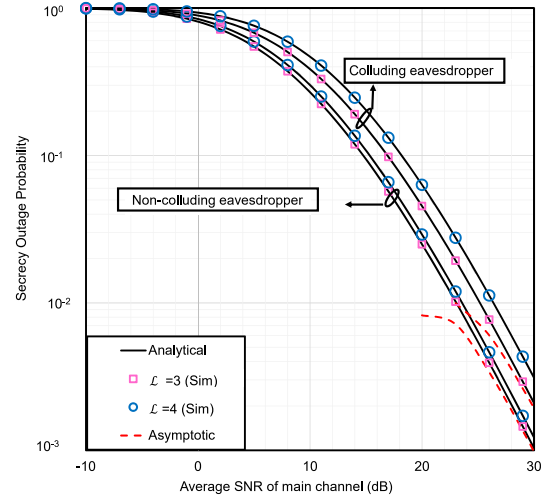}}
    \vspace{0mm}
    \caption{SOP vs $\bar{\gamma}_d$ for the selected values of $\mathcal{L}$.}
  \label{r4}  
  \vspace{-4mm}
\end{figure}

%%%%%%%%%%%%%%%%%%%%%%%%%%%%%%%%%%%%%%%%%%%%
Figure \ref{r4} depicts the impact of the number of eavesdroppers along with the multiple eavesdroppers' collusion and non-collusion.
As can be observed, the SOP rises with an increase in $\mathcal{L}$. The impact of colluding eavesdroppers is observed as more harmful since they are able to collaborate and combine their resources and abilities to intercept confidential information.

% \begin{figure}[!ht]
% \begin{subfigure}{0.48\textwidth}
%     \includegraphics[width=\textwidth]{Figure/r2.eps}
%     \vspace{0.00mm}
%     \caption{Scenario I}
%     \label{r2}
% \end{subfigure}
% \begin{subfigure}{0.48\textwidth}
%     \includegraphics[width=\textwidth]{Figure/r3.eps}
%     \vspace{0.00mm}
%     \caption{Scenario II}
%      \label{r3}
% \end{subfigure}
% \caption{??????}
% \end{figure}

%%%%%%%%%%%%%%%%<SECTION>%%%%%%%%%%%%%%%
\section{Conclusions}\label{conclusion}
This study focuses on the secrecy outage performance of a RIS-aided wireless system in the presence of multiple eavesdroppers and interferers. All the results are justified by Monte-Carlo simulations and asymptotic analysis is also carried out at high SNR regime. The numerical results indicate that the impact of colluding eavesdroppers is more severe than that of non-colluding eavesdroppers. Moreover, the presence of interferers also imposes a detrimental effect on SOP, which can be reduced by adjusting the number of reflecting elements of both RIS.

%%%%%%%%%%%%%%%%<SECTION>%%%%%%%%%%%%%%%
\bibliographystyle{IEEEtran}
\bibliography{IEEEabrv,asmbBiblio.bib}

\end{document}